\documentstyle[preprint,prb,aps]{revtex}

\tighten

\begin{document}

\draft

\preprint{MA/UC3M/09/94}

\title{Three-dimensional effects on extended states in disordered
models of polymers}

\author{Francisco Dom\'{\i}nguez-Adame}

\address{Departamento de F\'{\i}sica de Materiales,
Facultad de F\'{\i}sicas, Universidad Complutense,\\
E-28040 Madrid, Spain}

\author{Enrique Diez and Angel S\'anchez}

\address{Escuela Polit\'ecnica Superior,
Universidad Carlos III de Madrid,
C./ Butarque 15, \\ E-28911 Legan\'es, Madrid, Spain}

\date{\today}

\maketitle

\begin{abstract}

We study electronic transport properties of disordered polymers in a
quasi-one-dimensional model with fully three dimensional interaction
potentials.  We consider such quasi-one-dimensional lattices in the
presence of both uncorrelated and short-range correlated impurities.  In
our procedure, the actual physical potential acting upon the electrons
is replaced by a set of nonlocal separable potentials, leading to a
Schr\"odinger equation that is exactly solvable in the momentum
representation.  By choosing an appropriate potential with the same
spectral structure as the physical one, we obtain a discrete set of
algebraic equations that can be mapped onto a tight-binding-like
equation.  We then show that the reflection coefficient of a pair of
impurities placed at neighboring sites (dimer defect) vanishes for a
particular resonant energy.  When there is a finite number of such
defects randomly distributed over the whole lattice, we find that the
transmission coefficient is almost unity for states close to the
resonant energy, and that those states present a very large localization
length.  Multifractal analysis techniques applied to very long systems
demonstrate that these states are truly extended in the thermodynamic
limit.  These results are obtained with parameters taken from actual
physical systems such as polyacetylene, and thus reinforce the
possibility to verify experimentally theoretical predictions about
absence of localization in quasi-one-dimensional disordered systems.

\end{abstract}


\narrowtext

\section{Introduction}

Transport properties of disordered systems have become a fascinating
research topic since the generality of localization phenomena in one
dimension (1D) was first questioned a few years ago.
\cite{Flores,Dunlap,Wu,Wu2,Wu3,Bovier,Wu4,Evan2,indios,Evan1,Flores2,%
nozotro,PRBKP,JPA,Evan3,Datta,Diez,Brito} Opposite to the conventional
view that in 1D random systems almost all eigenstates are exponentially
localized (see, e.\ g., Ref.~\onlinecite{Ziman} and references therein),
it is nowadays known that in disordered systems where disorder exhibits
some kind of spatial correlation bands of extended states arise.
Spatial correlation means that random variables are not independent
within a given correlation length or, equivalently, that the noise is
non-white.  Furthermore, supression of localization by structural
correlation has been found both in classical and quantum systems.  In
the quantum case, electronic transport has been of course the subject of
most works.  There exists at present much evidence that correlated
disorder inhibits wave localization, and that bands of extended states
appear in tight-binding Hamiltonians \cite{Flores,Dunlap,Wu,Wu2,Wu3,%
Bovier,Wu4,indios,Evan1,Flores2} as well as in more elaborated multiband
systems like those described by Kronig-Penney models.\cite{PRBKP,JPA}
Similarly, occurrence of super-diffusion\cite{Evan2} and reflectionless
spin waves in Heisenberg chains\cite{Evan3} have been recently reported.
In the classical case, random harmonic chains also present a band of
shortwave delocalized vibrations whenever correlated disorder occurs,
\cite{nozotro,Datta} giving rise to a strong enhancement of the thermal
conductivity of the lattice.\cite{Brito} All these theoretical analyses
clearly demonstrate that transport properties in random systems where
structural correlations are present are very different to what is found in
ordinary random systems.  It is also clear that supression of
localization does not depend on the classical or quantum nature of the
system and therefore structural correlations are to be regarded as the
origin of this unexpected feature.

In spite of the already available body of theoretical work, the physical
relevance of these extended states is still unknown.  To our knowledge,
there is no experimental evidence whatsoever of the existence of these
states and their influence in measurable transport properties.  We
regard this as the key question to be posed about the theoretical
results, above the more fundamental one on the nature of these states in
infinite systems.  Notwithstanding, we have addressed both issues along
our research project on disordered systems \cite{nozotro,PRBKP,JPA,%
Diez,Brito,exciton,optical,exciton2} and in particular in the work we
are reporting here.  Regarding experimental demonstration of
delocalization, we have recently shown how the bands of extended states
must reveal themselves through characteristic features in the dc
conductance of disordered superlattices at finite temperature.
\cite{Diez} Moreover, we have also found that short-range correlated
disorder has profound effects on coherent\cite{exciton} and incoherent
\cite{exciton2} trapping, as well as on optical properties
\cite{optical} of excitons.  In the same spirit, we have even proposed
mechanical analogues where classical extended vibrations should be
found. \cite{Brito} We note that, aside from the basic research goal of
finding out whether delocalization actually occurs in real physical
systems, we also have in mind an applied aim, namely verifying whether
correlated disorder gives rise to particular features that can be used
for new devices or applications.

In this work, we report further progress along the lines in the
preceding paragraph.  Searching for physically realizable systems where
delocalization may play a crucial r\^ole, we turn ourselves to one of
the pioneering works in the field, namely the work of Wu and Phillips on
polyaniline (see Ref.~\onlinecite{Wu3} and references therein).  These
authors showed that polyaniline could be mapped onto a tight-binding,
random dimer model that has a band of extended states, originated by a
resonance of a single dimer defect.  It is evident that on the one hand,
similar mappings can be worked out for different polymers, and on the
other hand, that delocalized bands have to affect their conductance
properties.  Indeed, Wu and Phillips argued that the fact that
polyaniline was a conducting polymer was closely related to this
unexpected delocalization phenomena.\cite{Wu2} Their calculations were
carried out in the framework of a purely 1D tight-binding Hamiltonian.
However, although it is quite reasonable to approximate the structure of
a polymer by a line, it is also true that the physics involved is
three-dimensional (3D) and that actually the linear structure of
polymers is not straight but folds and wanders in 3D space.  It is then
natural to ask whether the above theoretical results will still hold
when more realistic models including 3D effects are considered.  The
answer to this question is very important if Wu and Phillips's results
are to be compared to measurements on real polymers: If the delocalized
band is destroyed by 3D effects, then their theoretical results are
merely academic, and worse, any possible technological application
becomes very unlikely.  On the other hand, this question is not without
interest from the fundamental viewpoint.  As we mentioned above,
localization of almost all eigenstates by uncorrelated disorder is
expected in 1D random systems, but three-dimensional (3D) systems
require a minimum amount of disorder to give rise to localization.
\cite{Ziman} Then a question arises in a natural way within this
context, namely the possible effects of correlated disorder on 3D
eigenstates.  As far as we know, this problem has been already studied
by Stephens and Skinner,\cite{Stephens} who found that tight-binding
Hamiltonians with short-range correlated diagonal disorder in a cubic
lattice present a localization threshold that is independent of the
amount of correlation.  This finding seems to indicate that the
influence of structural correlations is relevant only to pure 1D
systems, which adds further interest to the elucidation of the
applicability of Wu and Phillips's results.

We address the above issues by introducing a new and completely general
model to study electronic properties in random systems based on the
so-called nonlocal (separable) potential (NLP) method, in which the
actual potential at each site of an arbitrary lattice is replaced by a
projective operator. \cite{Knight,Sievert,MP} The treatment is fully 3D,
although we restrict ourselves to a linear chain, so our model is {\em
not} a 1D model in the traditional sense as we have here an array of 3D
potentials along a straight line.  Moreover, the model can be
straightforwardly extended to folded (i.e., non-straight) systems by
appropriate choices of the parameters.  As a major point, we will
demonstrate the occurrence of a well-defined band of extended states in
the electronic energy spectrum due to structural correlations, in spite
of the 3D character of the equation of motion.  In particular, we
consider the case in which pairs of impurities (the so called {\em dimer
defects}) are placed at random in an otherwise perfect lattice.  The
location of the band of the extended states in the electronic spectrum
is determined from the condition of vanishing of the reflection
coefficient from a single dimer defect.

The paper is organized as follows.  In Sec.~II we present our model and
summarize previous work of us\cite{MP} that is necessary for a better
understanding of the present paper.  After describing how nonlocal
potentials can be used to model 3D systems, we consider the scattering
from a single dimer defect exactly and find the resonance condition for
perfect transparency.  We close this section with a brief account of
exact expressions to compute the physical magnitudes of interest in a
lattice containing a certain number of dimer defects.  Afterwards, in
Sec.~III we concern ourselves with our main topic, the random
quasi-one-dimensional lattice with paired disorder.  We present our
numerical results demonstrating the existence of extended states via
transmission and Lyapunov coefficients, which relate to physically
relevant quantities such as localization length, as well as multifractal
analysis, which points out the character of these states in the infinite
size limit: Thus, we show how the bands of extended states reveal
themselves through well defined peaks in the transmission coefficient
versus energy plots, and how the truly extended character of those
states is also demonstrated in their scaling properties.  Section IV
concludes the paper with a brief summary of the main results and some
possible applications in different physical contexts.

\section{NLP approach to multicenter interactions}

The starting point for the NLP procedure is the Schr\"odinger equation
for multicenter potentials, corresponding to the physical situation we
want to gain insight into.  The solution of this kind of problem is of
widespread interest not only in condensed matter physics but also in
atomic or molecular physics.  As is well known, such solution is
expected to involve enormous intricacies since in most cases
prohibitively cumbersome calculations are needed.  Several methods have
been developed to study the motion of electrons in a given superposition
of 3D potentials.  Among them, the NLP approach is the natural
generalization of the famous Kronig-Penney model\cite{KP} to the 3D
case.  This method leads to an exactly solvable Schr\"odinger equation
from which the electron energy can be obtained in a closed form.  What
is more important, it is always possible to find a NLP (or a sum of
them) which reproduces any set of given electronic states,\cite{Larry}
so there is no theoretical limit to the numerical accuracy with which
physical results may be obtained.  We first summarize the NLP formalism
and then discuss its application to conducting polymers.

\subsection{Schr\"odinger equation for NLP}

We begin from the Schr\"odinger equation for NLP, which reads as
follows\cite{MP} (we take $\hbar=m=1$ hereafter)
\begin{equation}
\left( {\bf p}^2-2E \right) \psi ({\bf r}) =
\sum_k\> \lambda_k V(|{\bf r}-{\bf R}_k|) \int \> d^3r'
V(|{\bf r}'-{\bf R}_k|) \psi ({\bf r}' ),
\label{1}
\end{equation}
where ${\bf R}_k$ denotes the position of each lattice site and
$\lambda_k$ is the corresponding coupling constant. We will immediately
see how Eq.~(\ref{1}) connects with the physical problem of interest,
through suitable choices of the potential $V$.  For simplicity we have
assumed that the function $V$ is spherically symmetric, although more
complicated symmetries can be also easily handled.  In Fourier space we
have
\begin{equation}
\psi ({\bf p}) = \left( {1\over {\bf p}^2-2E} \right)
\sum_k\> \lambda_k V(p)\exp (-i{\bf p}\cdot{\bf R}_k) \chi_k,
\label{2}
\end{equation}
where
\begin{equation}
\chi_k=\int \>d^3p\,V^*(p)\exp(i{\bf p}\cdot{\bf R}_k)\psi({\bf p}).
\label{3}
\end{equation}
Here $\psi({\bf p})$ and $V(p)$ denote the Fourier transforms of $\psi$
and $V$, respectively.  The asterisk means complex conjugation; the
Fourier transform of real and spherically symmetric functions is also
real, but we retain it for if non-spherical functions are to be considered.
The coefficients $\chi_k$ are the quantities of interest, since we will
show that they are related to the wave function in real space.  We will
be more specific about their meaning after we have specified the
potential $V(p)$ and computed the corresponding equations for $\chi_k$.
Inserting Eq.~(\ref{2}) in Eq.~(\ref{3}) we obtain the following set
of algebraic equations for the parameters $\chi_k$
\begin{equation}
\chi_k=\sum_j\> \lambda_j \int \>d^3p\,\frac{|V(p)|^2}{{\bf p}^2-2E}
\exp [i{\bf p}\cdot({\bf R}_k-{\bf R}_j)]\chi_j.
\label{4}
\end{equation}
Due to the spherical symmetry of the potential, the angular integration
can be carried out in Eq.~(\ref{4}).  In so doing, we finally obtain
\begin{equation}
\chi_k=\sum_j\> 4\pi\lambda_j\, \int_0^\infty \>dp\,
\frac{p^2|V(p)|^2}{{\bf p}^2-2E}\
\frac{\sin pR_{kj}}{pR_{kj}}\chi_j,
\label{5}
\end{equation}
where $R_{kj}\equiv |{\bf R}_k-{\bf R}_j|$. It is understood that the
factor $(\sin pR_{kj})/pR_{kj}$ is replaced by $1$ when $k=j$, that is, by
its limit as $R_{kj}\to 0$.

\subsection{Application to quasi-one-dimensional polymer models}

At this point we should stress that Eq.~(\ref{5}) is completely general,
once the potential $V(p)$ is specified.  For a given 1D, 2D or 3D
lattice $\{ {\bf R}_k, k=1, 2, \ldots N\}$, $N$ being the number of
sites, the eigenenergies can be found by solving the secular equation
arising from the $N \times N$ symmetric determinant associated to
Eq.~(\ref{5}).  In this fashion, we arrive at the key of the NLP
procedure: The crucial question is to set up an appropriate potential
$V(p)$ that reproduces the observed energy values of the physical
situation being considered.  For instance, we have previously found that
the Yamaguchi's NLP \cite{Yama} is most appropriate to describe Coulomb
bound states (see Ref.~\onlinecite{MP}), whereas surface
$\delta$-function potentials, that is, a force field vanishing
everywhere except on a spherical shell of radius $R$, are very well
suited to simulate electron potentials in long quasi-one-dimensional
polymers, as polyacetylene or polyaniline.\cite{MP} This is the case
we are interested in, and therefore we concentrate ourselves in this
choice of potential from now on, i.\ e., we take
\begin{equation}
V(r)={1\over r^2}\,\delta (r-R), \hspace{7mm}
V(p)=\sqrt{2\over \pi} \, {\sin pR\over pR}.
\label{6}
\end{equation}
Plugging this potential into Eq.~(\ref{5}) we get
\begin{equation}
\chi_k= 8\,{\lambda_k\over R^2}\, A(E)\chi_k + \sum_{j\neq k}\>
8\,{\lambda_j\over R^2}\, B_{kj}(E)\chi_j,
\label{7}
\end{equation}
where for brevity we have defined
\begin{eqnarray}
A(E) & = & \int_0^\infty \>dp\, \frac{\sin^2 pR}{{\bf p}^2-2E}
= {\pi \over 4\kappa} \left(1-e^{-2\kappa R}\right) \nonumber \\
B_{kj}(E) & = & \int_0^\infty \>dp\,
\frac{\sin^2 pR}{{\bf p}^2-2E}\ \frac{\sin pR_{kj}}{pR_{kj}}=
{\pi \exp (-\kappa R_{kj}) \over 4\kappa^2R_{kj}}
\left( \cosh 2\kappa R-1\right).
\label{8}
\end{eqnarray}
We are restricting ourselves to the case of interest, namely $E<0$ and
then $\kappa\equiv \sqrt{-2E}$ is a real parameter.

We note that interference effects due to the interaction of the electron
with the lattice appear in the coefficients $B_{kj}(E)$: The larger the
distance between site $k$ and $j$, the smaller the corresponding
coefficient.  In other words, such coefficients are rapidly decreasing
functions of $R_{kj}$ whenever $\kappa$ is not very small (deep
potentials).  In our problem, this is a good approximation, and hence,
to simplify numerical analysis, we assume that only nearest-neighbor
interactions along the linear lattice are significant and write
\begin{equation}
\chi_k= 8\,{\lambda_k\over R^2}\, A(E)\chi_k +
8\,{\lambda_{k+1}\over R^2}\, B_{k\,k+1}(E)\chi_{k+1} +
8\,{\lambda_{k-1}\over R^2}\, B_{k\,k-1}(E)\chi_{k-1}.
\label{9}
\end{equation}
To evaluate the coupling constant $\lambda_k$ in terms of experimentally
measurable quantities, we consider an isolated potential centered at
${\bf R}_k$.  This we accomplish by neglecting the interaction with
other lattices sites so we take $B_{k\,k\pm1}(E)\to 0$ in Eq.~(\ref{9}),
thus obtaining the condition determining the energy of bound states
$E_k$, namely $8\lambda_k A(E_k)/R^2=1$.  This is a transcendental
equation which can be easily solved numerically, but it leads to a
simpler expression for small values of $R$, a limiting case we will
consider later.  For small radius $R$ we can expand $A(E_k)$ to obtain a
relationship between the coupling constant $\lambda_k$ and the energy of
the (single) bound state $E_k=-\kappa_k^2/2$

\begin{equation}
\lambda_k \sim {R\over 4\pi}(1+\kappa_k R).
\label{10}
\end{equation}
Finally, inserting Eq.~(\ref{10}) in Eq.~(\ref{9}) and taking the limit
$R\to 0$ in such a way that $\kappa_k$ remains constant, we obtain the
following tight-binding-like equation for the coefficients $\chi_k$
\begin{equation}
(\kappa-\kappa_k)\chi_k=
\frac{\exp (-\kappa R_{k\,k+1})}{R_{k\,k+1}}\chi_{k+1}+
\frac{\exp (-\kappa R_{k\,k-1})}{R_{k\,k-1}}\chi_{k-1}.
\label{11}
\end{equation}
Note that in this tight-binding set of equations the transfer integrals
depend exponentially on the distance between nearest-neighbor distance,
as it should be expected.  This is consistent with our previous
disregarding of longer range interactions.  We can use these equations
to describe the dynamics of electrons in the presence of diagonal as
well as off-diagonal randomness.  In the rest of the paper, and without
loss of validity, we further assume diagonal disorder, which implies
that $R_{k\,k\pm 1}=L$, $L$ being the lattice parameter.  Defining
$\rho\equiv \kappa L$ and $\rho_k\equiv \kappa_k L$ for the sake of
brevity, we thus arrive at
\begin{equation}
(\rho-\rho_k)e^\rho\chi_k=\chi_{k\,k+1}+\chi_{k\,k-1}.
\label{12}
\end{equation}
It is most important to stress that the number of free parameters
appearing in these equations of motion has been kept to a minimum: We
have only introduced the strength of the potential (which manifests
itself in the value of the single bound state level, appearing in
$\rho_k$) and the lattice parameter $L$.

Before we proceed to study the above equations of motion for our 3D
model polymer, we now clarify the physical meaning of the coefficients
$\chi_k$.  From their definition in Eq.~(\ref{3}) and the Parseval
identity \cite{Gali} we have
\begin{equation}
\chi_k=\int \>d^3r\,V(r)\psi({\bf r}+{\bf R}_k),
\label{13}
\end{equation}
with $V(r)=(1/r^2)\delta (r-R)$. In the limit $R\to 0$ one gets
$V(r)\to (1/r^2)\delta (r)=\delta({\bf r})$. Therefore, in that limiting
case
\begin{equation}
\chi_k=\psi({\bf R}_k).
\label{14}
\end{equation}
We thus see that $\chi_k$ is nothing but the value of the electron
wave function at site ${\bf R}_k$, which is of course the quantity of
interest. On the other hand, this is the reason of the
denomination of Eq.~(\ref{12}) as {\em tight-binding} equations of
motion.

\subsection{Perfect quasi-one-dimensional lattices}

Before considering random lattices, it is instructive to study the case
of perfect lattices, that is, those lattices with $\rho_k=\rho_0$. Since there
exists translational symmetry, Bloch theorem holds and we look for
solutions of the form $\chi_k=U\exp(iQLk)$, $Q$ being the crystal
momentum and $U$ a constant. Inserting this solution in the Eq. (\ref{14})
we readily obtain the dispersion relation
\begin{equation}
\cos QL = {e^\rho\over 2}\, (\rho-\rho_0).
\label{15}
\end{equation}
Real values of $Q$, obtained by usual numerical methods, give us the
electron energy $E=-\rho^2/2L^2$ as a function of $Q$ and, consequently,
the band structure of the lattice.  To check the validity of the
tight-binding approach we have assumed, it is necessary to compare this
band structure with that obtained by including all non-nearest-neighbor
interactions.  We have already calculated it in Ref.~\onlinecite{MP},
obtaining
\begin{equation}
\cos QL = \cosh \rho-\,{1\over 2}e^{\rho_0}.
\label{16}
\end{equation}
Assuming that the lattice parameter $L$ is large and the potentials are
deep (the basic assumptions in the tight-binding approach) it becomes
clear that $\rho_0$ and $\rho$ are large but the difference
$\rho-\rho_0$ is small.  With this assumptions it is a matter of simple
algebra to demonstrate that Eq.~(\ref{16}) reduces to Eq.~(\ref{15}).
This leads us to the conclusion that one can confidently use
Eq.~(\ref{15}) to describe the motion of tightly bound electrons in a
lattice.

{}From an experimental point of view, comparison with measurements of
real polymers requires the evaluation of the two input parameters,
namely $L$ and $\rho_0$, from experimental data.  The first one is
usually known from X-ray data and, in principle, it is easy to obtain.
The second one requires more information of the electronic band
structure obtained, for instance, from spectroscopy measurements.  Let
us assume for the moment that the energies of the experimental band
edges are known, and let $E_t$ and $E_b$ be the energy of the top and
the bottom of the band in the perfect lattice, respectively ($QL=0$ and
$QL=\pi$), and $\rho_t=L\sqrt{-2E_t}$ and $\rho_b=L\sqrt{-2E_b}$.  Using
Eq.~(\ref{15}) one has the following relationship
\begin{equation}
\rho_0=\frac{\rho_t\exp(\rho_t)+\rho_b\exp(\rho_b)}{\exp(\rho_t)+
\exp(\rho_b)}
\label{17}
\end{equation}
Therefore, from the knowledge of the experimental band edges we can
calculate the semi-empirical parameter $\rho_0$. We have thus shown how
the model parameters can be found for comparison to the particular
polymer one is interested in. We will make use of this result later.

\subsection{Scattering from a single dimer defect}

As mentioned in the Introduction, we are interested in the effects of
structural correlations in the localization properties of
quasi-one-dimensional polymer models.  Following Wu and
Phillips,\cite{Wu2} the simplest way to consider structural correlations
is to introduce impurities at random but in pairs of sites.  Physically
this would correspond, for instance, to complexes of defects frequently
encountered not only in polymers but also in molecular and solid state
physics.  In particular, a very clear description of dimer defects in
polyaniline can be found in Ref.~\onlinecite{Wu3}.  In our model this
means that $\rho_k$ can take only two values, $\rho_0$ and $\rho_0'$,
with the additional constraint that $\rho_0'$ appears only in pairs of
neighboring sites, which we will refer to as dimer defects.

Let us consider a single dimer defect placed at sites $k=0$ and $k=1$ in
an otherwise perfect lattice. To proceed, we have to take into account
in the first place the condition for an electron to move in the perfect
lattice which, recalling Eq.~(\ref{15}), is given by
\begin{equation}
\left| {e^\rho\over 2}\, (\rho-\rho_0) \right| \leq 1;
\label{18}
\end{equation}
this constraint gives the allowed energy values once $\rho_0$ is fixed.
Now considering the equation of motion, Eq.~(\ref{12}), at $k=-1,0,1$
and eliminating $\chi_0$ and $\chi_1$ on gets
\begin{equation}
-\chi_2=(\Omega+\Omega'-\Omega\Omega'^2)\chi_{-1}-(1-\Omega'^2)\chi_{-2},
\label{19}
\end{equation}
where we have defined $\Omega\equiv e^\rho(\rho-\rho_0)$ and
$\Omega'\equiv e^\rho(\rho-\rho_0')$ for brevity.  Besides a constant
phase factor of $\pi$, Eq.~(\ref{19}) reduces to the equation of motion
in the perfect lattice whenever $\Omega'=0$, in which sites $k=0$ and
$k=1$ have been eliminated.  This means that the reflection coefficient
at the single dimer vanishes, and consequently there exists a complete
transparency.  This occurs only for a particular energy of the incoming
electron $E_r\equiv -\rho^2_r/2L^2$, given by the condition $\Omega'=0$,
i.e. $\rho_r=\rho_0'$.  Hence this resonance effect occurs whenever the
incoming electron matches the energy level of the (isolated) impurity,
and this is possible only if $|\rho_0-\rho_0'|\leq 2\exp(-\rho_0')$, as
seen from Eq.~(\ref{18}).  This is to be compared with both the results
of Ref.~\onlinecite{Wu2}, where a single resonant energy is found as
well, and to those in Refs.~\onlinecite{PRBKP} and \onlinecite{JPA}
for a continuum
Kronig-Penney random dimer model, where an infinite number of resonances
arise.  We see that delocalization effects of structural correlation may
be more or less dramatic depending on the physical situation studied.
On the other hand, the important result is that the resonance of the
simple 1D tight-binding random dimer model is preserved in our 3D setup.

\subsection{Scattering from a lattice with random dimer defects}

We now proceed to the problem of a random lattice with a finite number
of dimer defects.  Of course, the above results do not imply anything
about extended states in a lattice with a {\em finite} number of dimers
defects, and it is necessary to study that problem separately.  For
definiteness, we introduce the concentration of defects $c$ given by the
ratio between the total number of impurities (twice the number of dimer
defects) and the total number of sites $N$ in the lattice.  We introduce
this definition to facilitate direct comparison with results in
ordinary random lattices with the same number of impurities and thus
the same value of $c$, although in the latter case there are no
constraints on the random location of the impurities.  To study
transmission properties of electrons through the random lattice, we
place it between two semi-infinite perfect lattices.  Therefore we
introduce the reflection $r$ and transmission $t$ amplitudes through the
relationships
\begin{equation}
\chi_k = \left\{ \begin{array}{ll} e^{i\rho k}+re^{-i\rho k}, & k<1, \\
                                   te^{i\rho k}, & k>N.
                 \end{array} \right.
\label{20}
\end{equation}
To determine both amplitudes we use the well-known transfer-matrix
techniques (see, e.\ g., Ref.~\onlinecite{Pendry}). Thus we cast
Eq.~(\ref{12})
into the matrix form
\begin{equation}
\left( \begin{array}{c} \chi_{k+1} \\ \chi_{k} \end{array} \right) =
\left( \begin{array}{cc} \alpha_k & -1 \\ 0 & 1 \end{array} \right)
\left( \begin{array}{c} \chi_{k} \\ \chi_{k-1} \end{array} \right)
\equiv P_{k}
\left( \begin{array}{c} \chi_{k} \\ \chi_{k-1} \end{array} \right),
\label{21}
\end{equation}
where $\alpha_k=(\rho-\rho_k)e^\rho$. The transfer matrix of the whole
lattice is then found as
\begin{equation}
T(N)=\prod_{k=N}^1\>P_k,
\label{22}
\end{equation}
which relates the wave function at both edges of the lattice.  Using the
fact that $\det (T)=1$ we finally arrive at the following expression for
the transmission coefficient $\tau=|t|^2$
\begin{equation}
\tau = \tau (E) = {4 \sin^2 \rho \over D(E)},
\label{23}
\end{equation}
with
\begin{eqnarray}
D(E) &=& T_{11}^2+T_{12}^2 +T_{21}^2+ T_{22}^2
+ 2 \left( T_{11}T_{12} + T_{21}T_{22}
- T_{11}T_{21} - T_{12}T_{22} \right) \cos \rho \nonumber \\
&-& 2 \left( T_{11}T_{22}+T_{12}T_{21} \right) \cos^2 \rho + 2 \sin^2
\rho,
\label{24}
\end{eqnarray}
where we have dropped the explicit dependence on $N$ of the
transfer-matrix elements.  The transmission coefficient $\tau$ can be
recurssively computed from the matrix elements of $T(N)$; taking into
account the fact that $T(N)=P_N\, T(N-1)$ and $T(0)=P_0$ we find the
following recurrence relations involving only real parameters
\begin{eqnarray}
T_{11}(N) &=& \alpha_N T_{11}(N-1)- T_{11}(N-2),  \nonumber \\
T_{12}(N) &=& \alpha_N T_{12}(N-1)- T_{12}(N-2),  \nonumber \\
T_{21}(N) &=& T_{11}(N-1),  \nonumber  \\
T_{22}(N) &=& T_{12}(N-1), \hspace{1 true cm} N=2,3\, \cdots
\label{25}
\end{eqnarray}
with the initial conditions $T_{ij}(0)=\delta_{ij}$, $T_{11}(1)=
\alpha_1$, $T_{12}(1)=-1$, $T_{21}(1)=1$ and $T_{22}(1)=0$.

Other physically relevant magnitudes can be readily obtained from the
transfer-matrix $T(N)$. In particular the Lyapunov coefficient, which
represents the rate of the growth of the wave function, is nothing but
the inverse of the localization length. It can be computed
as (measured in units of $L^{-1}$) \cite{Borland}
\begin{equation}
\Gamma (E) = \left( {1\over N} \right)
\left( T_{11}^2+T_{12}^2 +T_{21}^2+ T_{22}^2 \right).
\label{26}
\end{equation}
Delocalization of electronic wave function is seen through the decrease of
this parameter.

The results we have obtained so far provide an exact, although
nonclosed, analytical description of any random lattice with correlated
as well as uncorrelated disorder.  With them, we can compute the
magnitudes we mentioned above.  All expressions are very simple and
suitable for an efficient numerical treatment for any specific cases.
We will now evaluate them for several interesting cases to describe the
relevant features of the model and the fingerprints of extended states.

\section{Results and discussions}

There are five parameters that can be varied in our model, namely the
lattice constant $L$, the strengths of the scatterers $\rho_0$ and
$\rho_0'$, the total number of scatterers $N$, and the defect
concentration $c$.  In order to find results as close as possible to
actual systems, we consider a quasi-one-dimensional polymer, as is
the case of polyacetilene (CH)$_x$, which has been the focus of most of
the experimental and theoretical works.\cite{Fink} In the perfect
lattice, taking an uniform carbon-carbon bond length of
$L=1.39\,$\AA,\cite{Grant} we have previously estimated\cite{MP} that
$\rho_0=1.466$, corresponding to an energy level of the isolated
potential of $-4.23\,$eV.  We can confidently take these values as
correct since the predicted effective mass is found to be $m^*=1.65$, in
excellent agreement with the experimental result $m^*=1.7\pm 0.1$.  As
an example, we will consider $\rho_0'=1.550$ implying an energy level of
the isolated impurity of $-4.73\,$eV.  Note that the condition
$|\rho_0-\rho_0'|\leq 2\exp(-\rho_0')$ holds, that is, the energy level
of the isolated impurity lies in the band of the perfect lattice.
Hence, according to our previous considerations, there exits complete
transparency at an incoming energy of $-4.73\,$eV if only a single dimer
defect is placed in the lattice.  Now we must elucidate what happens
close to this resonant energy when several dimer defects are placed at
random in the lattice, in comparison with lattices with the same number
of {\em unpaired} defects.  We used lattice sizes ranging from
$N=2\,000$ to $N=500\,000$ sites.  The largest of these systems are
physically unrealizable, but it is important to study theoretically
those systems to clearly elucidate the truly extended character of
states close to the resonant energy, as we will demonstrate in the rest
of the paper; the results for the smallest values are those directly
related to experiments.  Concerning the fraction of impurities, we only
present here values corresponding to low defect concentration ($c$
ranging from 0.1 to 0.3) because of their more physical relevance to
actual systems, but we should stress that the main conclusion of the
paper, namely the existence of truly extended states in
quasi-one-dimensional lattices with correlated disorder, is independent
of $c$.

\subsection{Transmission coefficient}

Since we are dealing with random lattices, we will need ensemble
averages to compute the transmission coefficient.  Some years ago, Sak
and Kramer\cite{Sak} pointed out that only its logarithm obeys the
central limit theorem, thus being the unique physically representative
magnitude of the electron transmission, rather than the transmission
coefficient itself or its inverse.  Therefore we have actually
computed $\exp \langle \log \tau (E) \rangle$, where $\langle \ldots \rangle$
means ensemble average. Nevertheless, in what follows we refer this
quantity simply as the transmission coefficient and denote it by
$\tau$, but it is understood that averages are carried out over the
logarithm.

An example of the behavior of $\tau$ around the resonant energy is shown
in Fig.~\ref{fig1}, for both paired as well as unpaired lattices, with
the same values of $c=0.1$ and size $N=2000$.  A careful inspection of
the figure clearly reveals that $\tau$ is at least two orders of
magnitude larger in paired lattices than in unpaired ones in the region
of interest.  In addition, and what is more apparent, $\tau$ is close to
unity around the resonant energy $-4.73\,$eV, hence indicating that
perfect transparency is preserved even when a finite number of dimer
defects are placed at random in the lattice.  This is a signature of the
existence of a band of extended states close to that energy.  We stress
that, in spite of the fact that the plot corresponds to an average over
$300$ realizations, the transmission coefficient for typical
realizations {\em always} behaves in the same manner, although plots are
noisier.  Thus, the only appreciable effect of averaging is to smooth
out some particular very narrow, realization-dependent peaks, keeping
the main common wide peak centered at the resonant energy.

We want to highlight that the width of the transmission peak is always
nonzero.  Hence, close to the resonant energy, there is an interval of
energies that also shows high transparency, similar to that of the
resonant energy (note that there is a difference of about three orders
of magnitude between the transmission coefficient of paired and unpaired
lattices in that interval).  The peak width depends on the concentration
of dimers: The larger the concentration, the narrower the peak, being
always of finite width as already stated.  Figure~\ref{fig2} shows
results for two different values of $c$ (0.1 and 0.3) and $N=2000$.  In
addition, the width depends also on the system size: The larger the
size, the narrower the peak, as shown in Fig.~\ref{fig2} for two values
of $N$ ($1000$ and $2000$) for $c=0.1$.  It is worth mentioning that
$\tau$ is always unity at the resonant energy, irrespective of the value
of $c$ or $N$.

\subsection{Lyapunov coefficient}

The fact that the around the resonant energy $\tau$ becomes close to
unity suggests the possibility that the localization length of those
states may be very large.  This is, in fact, what is deduced from the
analysis of the Lyapunov coefficient (recall that it is the inverse of
the localization length).  Results are plotted in Fig.~\ref{fig3} for
paired as well as unpaired random lattices with $N=2000$ and $c=0.1$.
The comparison between the results for the two kind of lattices is
actually dramatic.  First of all, we again observe that there exists a
difference of several orders of magnitude between the localization
length in both systems.  In addition, paired lattices reflect the fact
that a large number of states around the resonant energy present a very
large localization length (which manifests itself in a deep minimum of
$\Gamma$, with the Lyapunov coefficient taking values of the order of
the inverse of the system size in a nonzero width region), whereas there
is a monotonic dependence of the Lyapunov coefficient for unpaired
disorder.

\subsection{Multifractal analysis}

{}From the study of the transmission coefficient and the Lyapunov
coefficient we are led to the conclusion that there exists a large
number of electronic states that remain unscattered (or almost
unscattered) by dimer defects.  Such states are characterized by very
large localization lengths (conversely, very small Lyapunov
coefficients).  Nevertheless, this result does not necessarily imply
that those states are truly extended, namely states that cannot be
normalized in the thermodynamic limit.  Then we must search for a
different approach in order to elucidate the localized or extended
character of the eigenstates.  The characterization of the spatial
extent of the wave function to all length scales may be accomplished by
means of multifractal analysis.\cite{tedesco} Extended states span
homogeneously the whole lattice whereas localized states remain confined
in finite regions.  The amplitude distribution of the electronic states
can be characterized by the scaling with the system size of moments
associated to the measure defined in the system by us (in our case the
probability of finding the electron at a given point),
\begin{equation}
\mu_q(N)=\frac{\sum_{n=1}^{N}\>|\chi_n|^{2q}}
{\left(  \sum_{n=1}^{N}\>|\chi_n|^{2}\right)^q}
\label{27}
\end{equation}
Notice that the second moment $\mu_2(N)$ coincides with the inverse
participation ratio (IPR) as introduced, for instance, in
Ref.~\onlinecite{Canisius}.  The generalized dimensions $D_q$ are
determined via the scaling $\mu_q(N) \sim N^{-(q-1)D_q}$, for $q\neq 1$.
For localized states $D_q$ vanishes for all $q$ whereas $D_q$ equals
unity for states spreading uniformly.  In previous works
\cite{nozotro,PRBKP} we have proven that multifractal analysis is a
powerful tool to reveal the existence of truly extended states in 1D
random systems (phonons, electrons) with correlated disorder.  Hence,
we expect that similar characterization techniques also work well
in quasi-one-dimensional systems, as in the present case.

Let us start with the IPR. From its definition, it can be seen that
delocalized states are expected to present small values of the IPR, of
order of $1/N$, while localized states have much larger values.  In the
extreme case, when the electron is localized at a single site,
Eq.~(\ref{27}) implies that $\mu_q(N)=1$.  A typical situation is
presented in Fig.~\ref{fig4} for the same system parameters as in
Fig.~\ref{fig1}, using the initial conditions $\chi_0=0$ and $\chi_1=1$
to iterate the equation of motion given by Eq.~(\ref{12}) in order to
find $\chi_k$.  One can observe a wide, deep minimum of the IPR around the
resonant energy for the paired disordered lattices, whereas this minimum
is completely absent in the unpaired one.  It is important to mention
here that the value of the IPR at the minimum is independent of the
defect concentration $c$, and depends only on the system size $N$.  This
suggest that the exact number of dimer defects is immaterial as far as
the existence of such extended states is concerned.

A complete multifractal analysis requires to study the scaling of all
moments, defined by Eq.~(\ref{27}) with system size.  We have consider
such scaling for $q=2,3 \ldots 6$, and results are plotted in
Fig.~\ref{fig5} for a concentration of $c=0.1$ of dimer defects.  We
have observed that those moments scale very accurately as $\mu_q(N)\sim
N^{-(q-1)}$ for energies close to the resonant one, as illustrated in
Fig.~\ref{fig5} for $-4.70\,$eV (close but not exact the resonant
energy).  On the contrary, for more distant energies $\mu_q(N)$ follows
a power law for small systems but tends to a constant value for larger
ones, as seen in Fig.~\ref{fig5} for $-4.50\,$eV.  Therefore, according
to the above discussion, the generalized dimensions $D_q$ are exactly
unity, within numerical accuracy, for states close to the resonant
energy, thus indicating the truly extended character of such states, in
agreement with results obtained from the analysis of the IPR above.

\section{conclusions}

In this paper we have considered electron dynamics in
quasi-one-dimensional models of polymers with correlated disorder, and
we have compared our results to those obtained in systems with ordinary
(uncorrelated) disorder.  Our procedure based on NLP allows us to carry
out a fully three dimensional analysis of the model, with the scatterers
placed along a straight line.  It is important to realize that this
technique can be made exact so there are no theoretical limitations in
this approach.  In addition, the exact solution can be found for an
arbitrary NLP, as we actually demonstrated [see Eq.~(\ref{4})] by means
of the Fourier transform (which, in fact, is completely equivalent to
use a Green's function formalism).  As our selection for a suitable NLP
that can reproduce experimental data for polymers, we have used surface
$\delta$-function interactions with vanishing radius since, as we
previously shown,\cite{MP} this potential gives very accurate results in
the context of polymers.  Using this model, we have found that there
exists a resonant energy for which the reflection coefficient of a
single dimer defect vanishes, that is, there is perfect transparency.
Afterwards, we turned to the problem of electron scattering when several
of such defects are located at random in the lattice.  Results from the
evaluation of the transmission coefficient and Lyapunov coefficient (the
inverse of the localization length) strongly suggest that there exist
many states close to the resonant energy that remain unscattered, where
this is not the case when the constraint of pairing is relaxed.  To
demonstrate that such states are actually extended in nature, we have
used multifractal analysis, which confirms our claim.

We now stress the physical relevance of our results.  A key observation
is that the resonant energy value does no depend on the defect
concentration $c$.  Therefore, by modifying this concentration, we could
shift the Fermi level of the quasi-one-dimensional lattice to match the
resonance.  In this case, when the Fermi level reaches the resonant
energy, a large increase should be observed in the electrical
conductance peak.  In fact, we have recently demonstrated in 1D
Kronig-Penney models with correlated disorder that very noticeable peaks
in the finite temperature dc conductance appear as the Fermi level is
moved through the band of extended states.\cite{Diez} In a similar way,
one could expect such dramatic increase in more elaborated models, as it
is the case with the one we present here.  On the other hand, our model
supports the results previously found for simpler, pure 1D
ones.\cite{Wu3} It seems to us that this agreement makes very appealing
the idea of use polymeric systems to confirm experimentally the
existence of delocalized states in 1D model, for what we have shown is
that 3D effects do not destroy the coherence required for
those states to appear.  In fact, it is very tempting to relate all this
to the known fact that polyaniline shows a metal-insulator transition,
with the concentration of dopant acting as a control parameter (see
Ref.~\onlinecite{Ducker} and references therein).  Another important
consequence of this work is that other experimental procedures we have
proposed to find out whether delocalization can be measured or not, such
as disordered superlattices,\cite{Diez} are likely to be correct even if
3D effects have not been taken into account.  We hope that this result
encourages experimental work in the field of quasi-one-dimensional
disordered systems which could give status of physically relevant to the
theoretically predicted bands of delocalized states.

\acknowledgments

We thank collaboration and illuminating conversations with E.\ Maci\'a.
Work at Madrid is supported by UCM through project PR161/93-4811.  Work
at Legan\'es is supported by the DGICyT (Spain) through project
PB92-0248, and by the European Union Human Capital and Mobility
Programme through contract ERBCHRXCT930413.

\begin{figure}
\caption{Transmission coefficient as a function of the energy around the
resonant energy $-4.73\,$eV, for paired (upper curve) and unpaired
(lower curve) random lattice.  Shown are averages over 300 realizations.
Every realization consists of $N=2000$ scatterers and a fraction of
defect $c=0.1$.}
\label{fig1}
\end{figure}

\begin{figure} \caption{Transmission coefficient as a function of the
energy around the resonant energy $-4.73\,$eV for paired lattices with
(a) $N=1000$ and $c=0.1$, (b) $N=2000$ and $c=0.1$, and (c) $N=2000$ and
$c=0.3$.  Shown are averages over 300 realizations.}
\label{fig2}
\end{figure}

\begin{figure}
\caption{Lyapunov coefficient as a function of the energy around the
resonant energy $-4.73\,$eV, for paired (lower curve) and unpaired
(upper curve) random lattice.  Shown are averages over 300 realizations.
Every realization consists of $N=2000$ scatterers and a fraction of
defects $c=0.1$.}
\label{fig3}
\end{figure}

\begin{figure}
\caption{IPR ($\mu_2$) as a function of the energy around the resonant
energy $-4.73\,$eV, for paired (lower curve) and unpaired (upper curve)
random lattice, with the same parameters as in Fig.~1.}
\label{fig4}
\end{figure}

\begin{figure}
\caption{Scaling of moments $\mu_2$ to $\mu_6$ with the system size in
paired random lattices for an energy $-4.70\,$eV (solid lines), i.\ e.,
close to the resonant energy, and for an energy $-4.50\,$eV (dashed
lines), i.\ e., far from the resonant energy.  Defect concentration is
$c=0.1$.}
\label{fig5}
\end{figure}

\end{document}